\documentclass[slac_one]{revtex4}
\usepackage{graphicx}
\usepackage{fancyhdr}
\begin{document}

\title{Data-driven Estimations of $Z$, $W$ and Top Backgrounds} 

%

\author{Mark Hodgkinson on behalf on the ATLAS Collaboration}
\affiliation{University of Sheffield, Sheffield, UK}

\begin{abstract}
The Standard Model processes of $W$, $Z$ and top quark production in
association with jets constitute a major background to searches for
Supersymmetry at the LHC. We describe recent work performed in the ATLAS
Collaboration to estimate these backgrounds for a basic SUSY selection, and
we discuss methods to derive them from the first ATLAS data.
\end{abstract}

\maketitle

\thispagestyle{fancy}


\section{INTRODUCTION} 
If Supersymmetry (SUSY) is realised in nature, an important step to claiming discovery of it is to understand the Standard Model backgrounds. Monte Carlo (MC) predictions may not be good enough to achieve this. Therefore data-driven background estimations have been studied. We use the one and no lepton search modes in $R$-parity conserving SUSY models with primary squark or gluino production as an example where data-driven background estimations can be used. We have used several scenarios based on the mSUGRA SUSY models \cite{susyCSC-ref}, each one corresponding to a particular set of values for 5 parameters in the model. We denote these models SUx where x can run from 1 to 8.

\vspace{-0.7cm}
\section{ONE-LEPTON SEARCH MODE}
Using a standard set of SUSY one lepton search cuts, defined in \cite{susyCSC-ref}, an excess of SUSY events is seen (Figure 1).
\begin{figure}[!h]
\begin{center}
$\begin{array}{c@{\hspace{0.5in}}c}
\multicolumn{1}{l}{\mbox{\bf (a)}} &
        \multicolumn{1}{l}{\mbox{\bf (b)}} \\ [-0.53cm]
\hspace{-0.4in}
\includegraphics[width=7.cm,height=3.4cm]{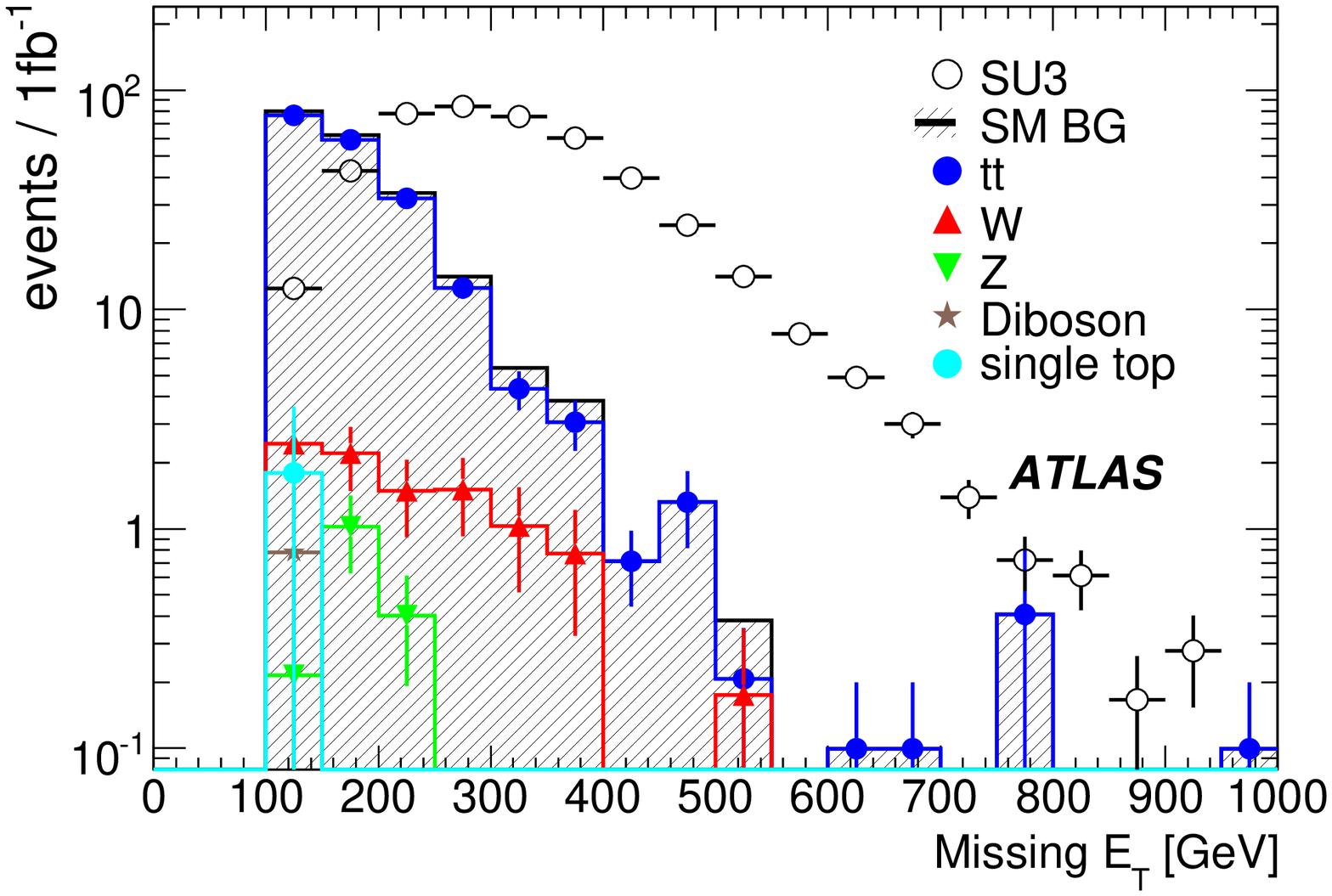} &
        \includegraphics[width=7.cm,height=3.4cm]{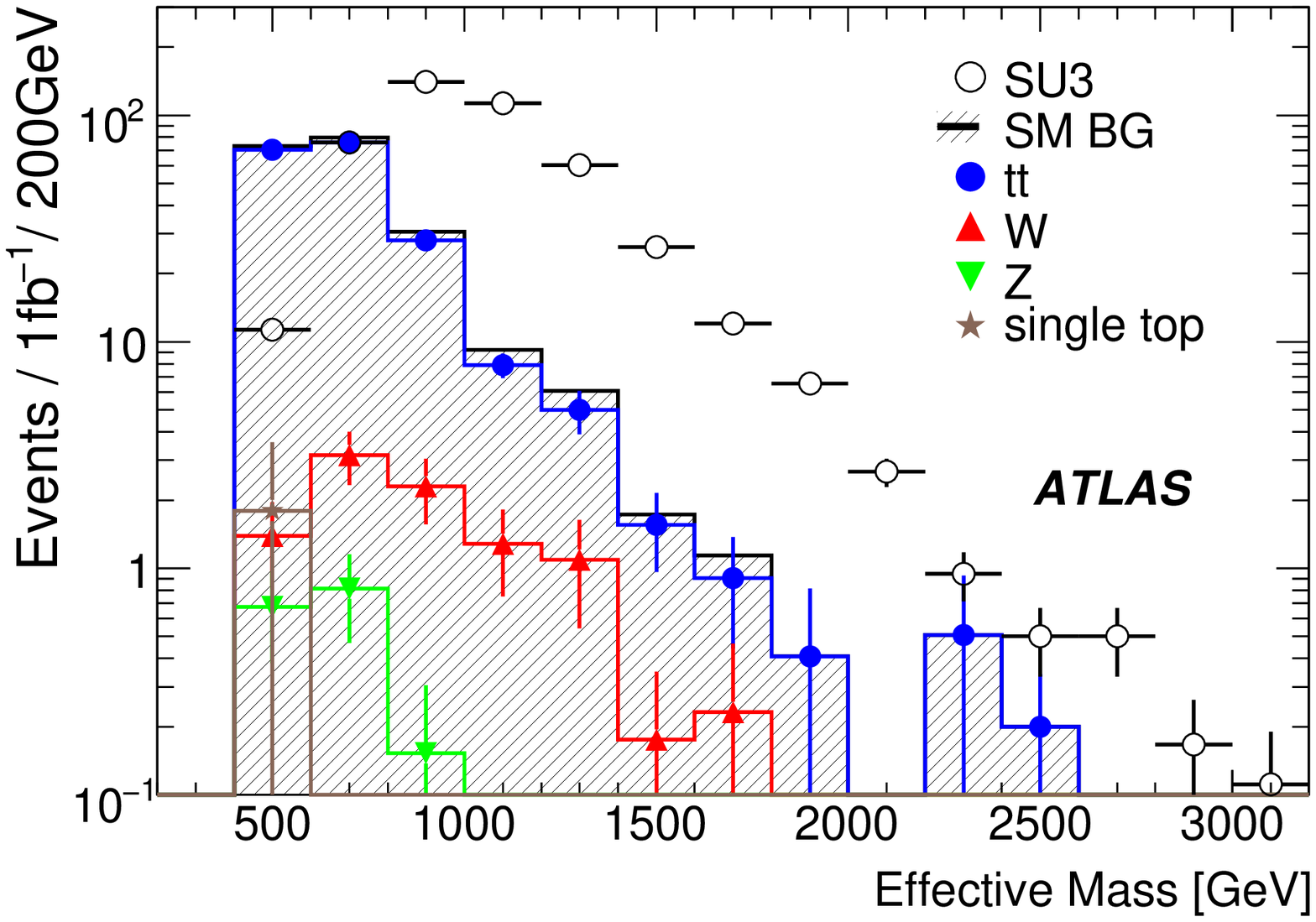} \\ [-0.4cm]
\end{array}$
\end{center}
\caption{For a SUSY model in the bulk region (denoted SU3 \cite{susyCSC-ref}) a clear excess of events would be seen in the missing transverse energy ($E_{T}^{miss}$) and effective mass distributions (scalar sum of the $P_{T}$ of the four hardest jets in the event with $|{\eta}|$ $<$ 2.5, all identified leptons and the $E_{T}^{miss}$) after the one lepton selection. The plots correspond to 1 fb$^{-1}$ of data. $W$ and top quark pair events dominate the background. }
\label{OneLepton_etmiss_meff}
\end{figure}
\vspace{-0.8cm}
\subsection{TRANSVERSE MASS ($M_{T}$) METHOD}
A control region, in which the shape of the $E_{T}^{miss}$ distribution for backgrounds can be estimated, is defined for this method for events with $M_{T}$ $<$ 100 GeV. The top quark pair (84$\%$) and $W$+jets (16$\%$) events are enhanced over SU3 in this region. In the absence of SUSY signal (left plot in Figure 2) the shape is correctly predicted in the signal region. However when SUSY signal is added to the data sample this is not the case (right plot in Figure 2). The systematic uncertainties on the predicted number of background events are up to 5$\%$ due to the jet energy scale, 7$\%$ due to lepton identification (ID) efficiency, 8$\%$ due to differences between the MC@NLO \cite{MCNLO-ref}\cite{MCNLO-refB} and ALPGEN \cite{ALPGEN-ref} generators and 5$\%$ due to MC parameter variation in ALPGEN.
\begin{figure}[h]
\begin{center}
\includegraphics[width=7.cm, height=3.5cm]{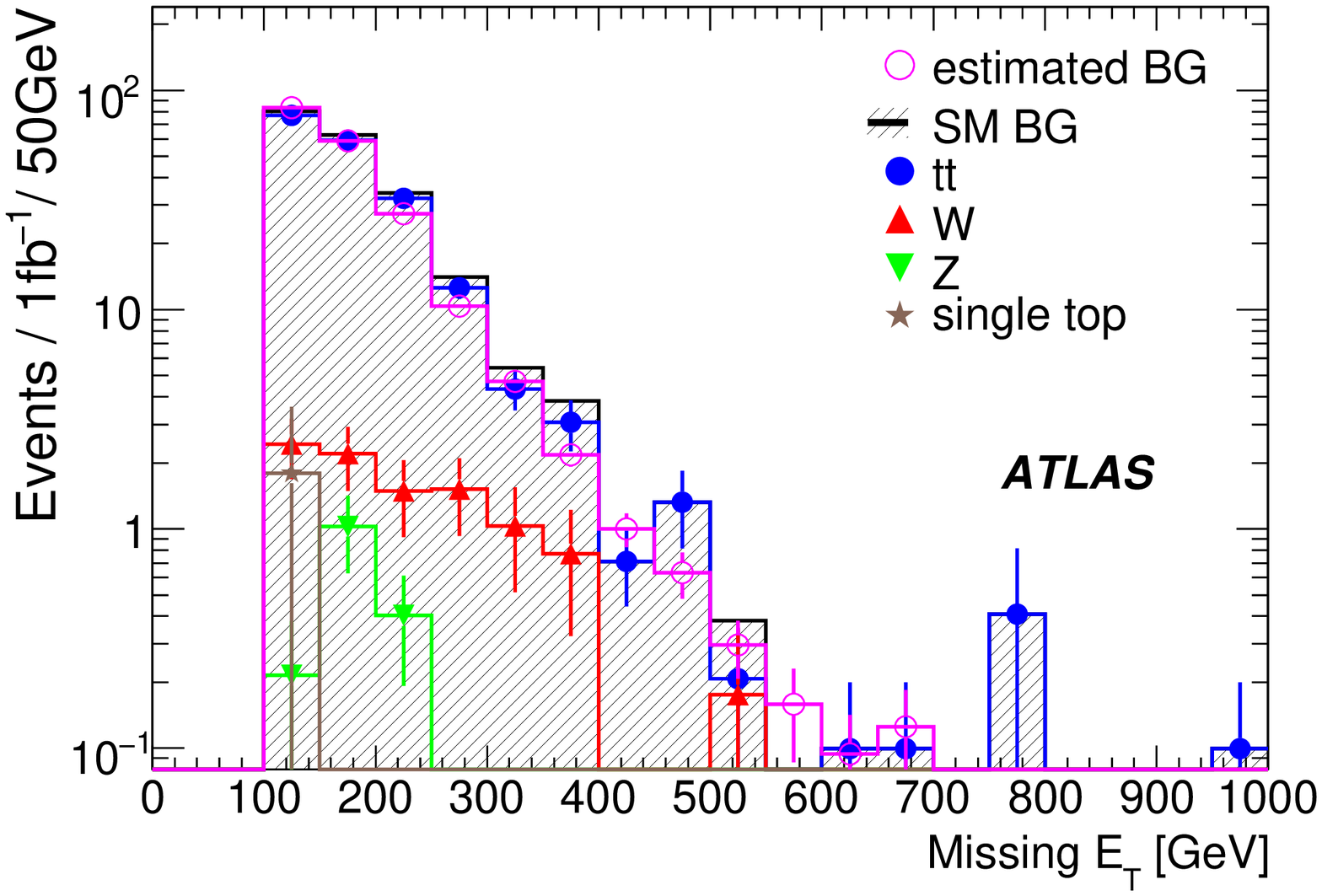}
\includegraphics[width=7.cm, height=3.5cm]{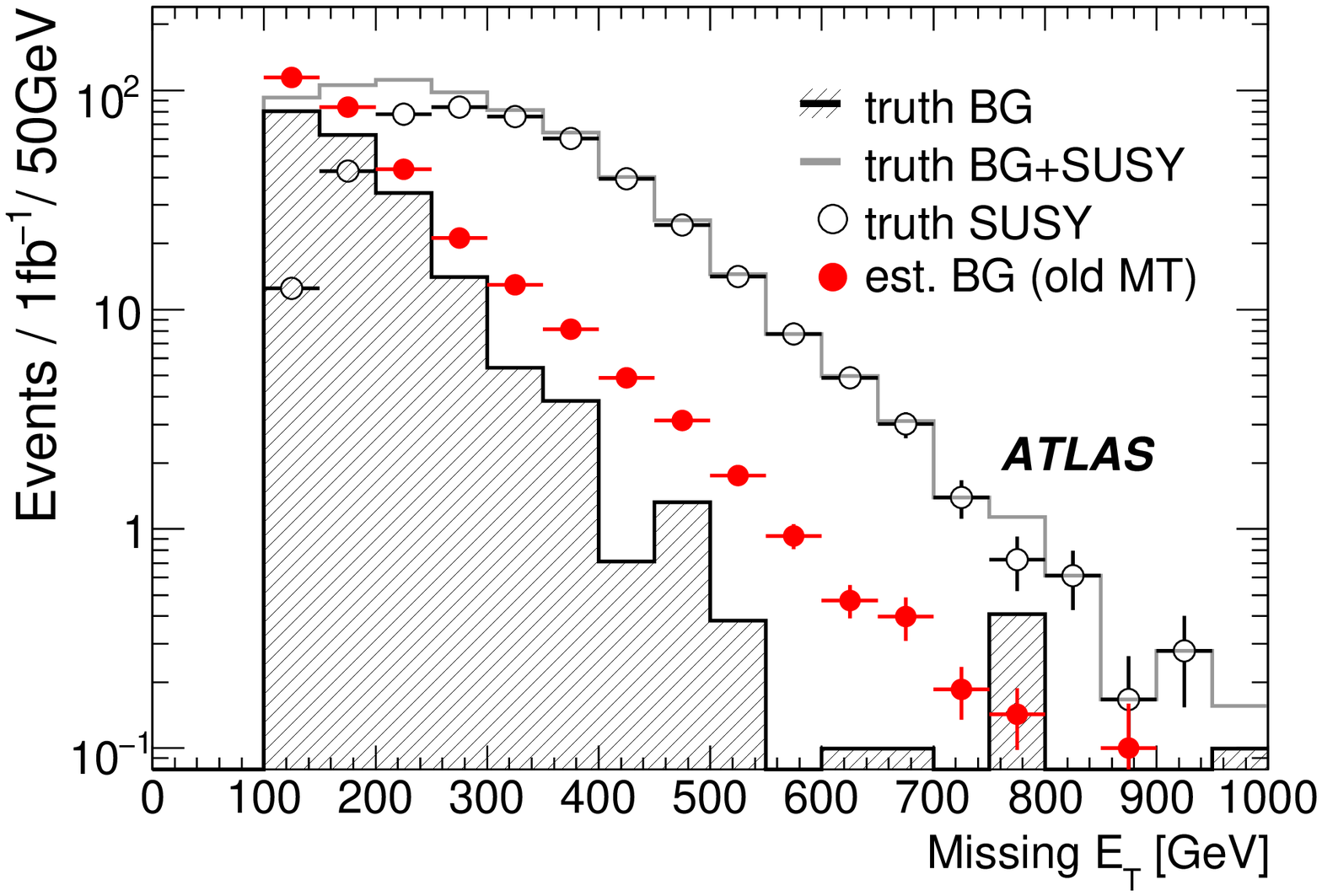}
\end{center}
\vspace{-0.5cm}
\caption{A comparison of the estimated background $E_{T}^{miss}$ and actual background $E_{T}^{miss}$ distributions in the absence of SUSY signal (left) and in the presence of SUSY signal (right). }
\end{figure}

For SU1 \cite{susyCSC-ref}, SU2 \cite{susyCSC-ref} and SU3 SUSY models the transverse mass falls slowly compared to Standard Model backgrounds (left plot in Figure 3). Therefore a general ansatz for the SUSY $M_{T}$ shape can be used to subtract the SUSY signal from the control region. The predicted $E_{T}^{miss}$ distribution for backgrounds then agrees with the actual background $E_{T}^{miss}$ distribution (right plot in Figure 3). Additional systematic uncertainties due to this technique, called the New $M_{T}$ technique, have not been evaluated currently.

\begin{figure}[h]
\begin{center}
\includegraphics[width=7.cm,height=3.5cm]{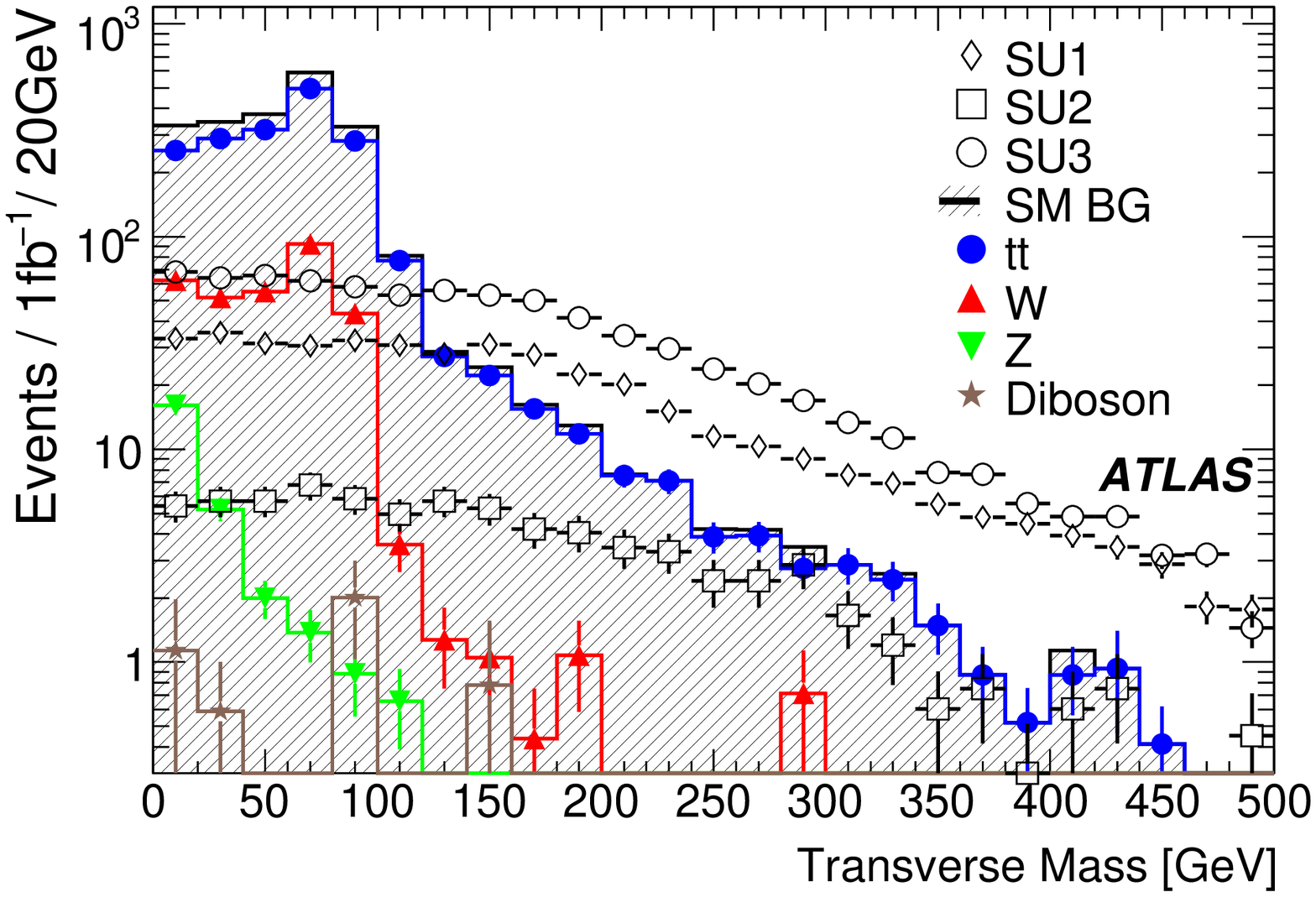} 
\includegraphics[width=7.cm,height=3.5cm]{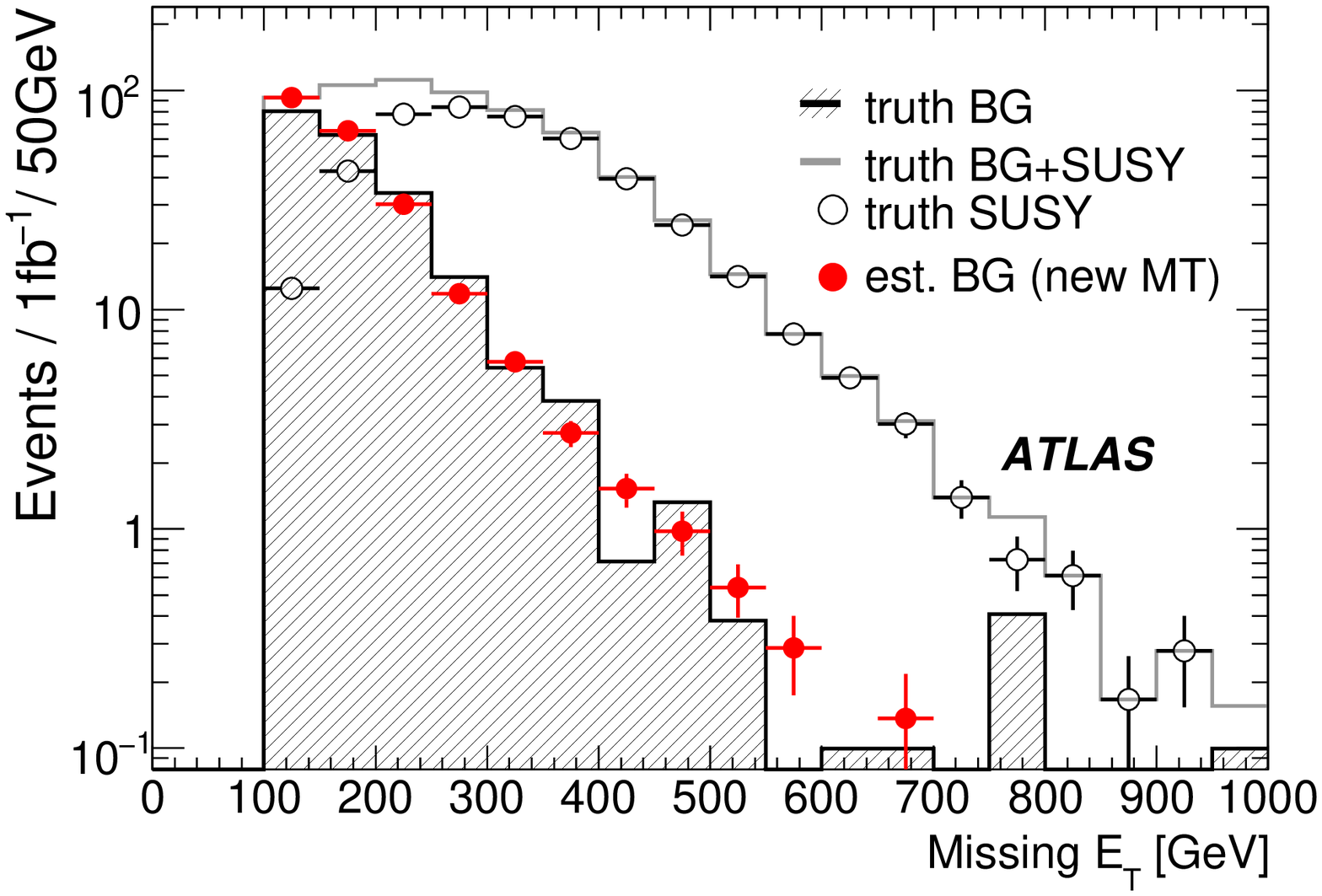} 
\end{center}
\vspace{-0.5cm}
\caption{Transverse mass distribution for SUSY signal and Standard Model background (left). Predicted and actual background $E_{T}^{miss}$ distributions using the New $M_{T}$ method. }
\end{figure}

\vspace{-1cm}
\subsection{DECAY RESIMULATION TECHNIQUE}
If one lepton is not identified dileptonic top events can be selected in the one lepton search. Therefore a control sample of fully leptonic top quark pair events is selected with cuts described in \cite{susyCSC-ref}. The top kinematics are reconstructed in this data control sample, the decay products removed and replaced with redecayed particles from PYTHIA 6.4 \cite{Pythia-ref}. The tops are decayed 1000 times each, making the approximation they are independent of one another, and the ATLAS fast simulation used to simulate detector effects. The left plot in Figure 4 shows good agreement in the $E_{T}^{miss}$ distribution for predicted and actual background at $E_{T}^{miss}$ $>$ 200 GeV. The number of background events with $E_{T}^{miss}$ $>$ 200 GeV can be predicted with a precision of 30$\%$ using this technique. If SUSY signal is added into the seed events the estimate is not as accurate, although an excess of SUSY is still seen.


\begin{figure}[h]
\begin{center}
\includegraphics[width=7.cm,height=3.5cm]{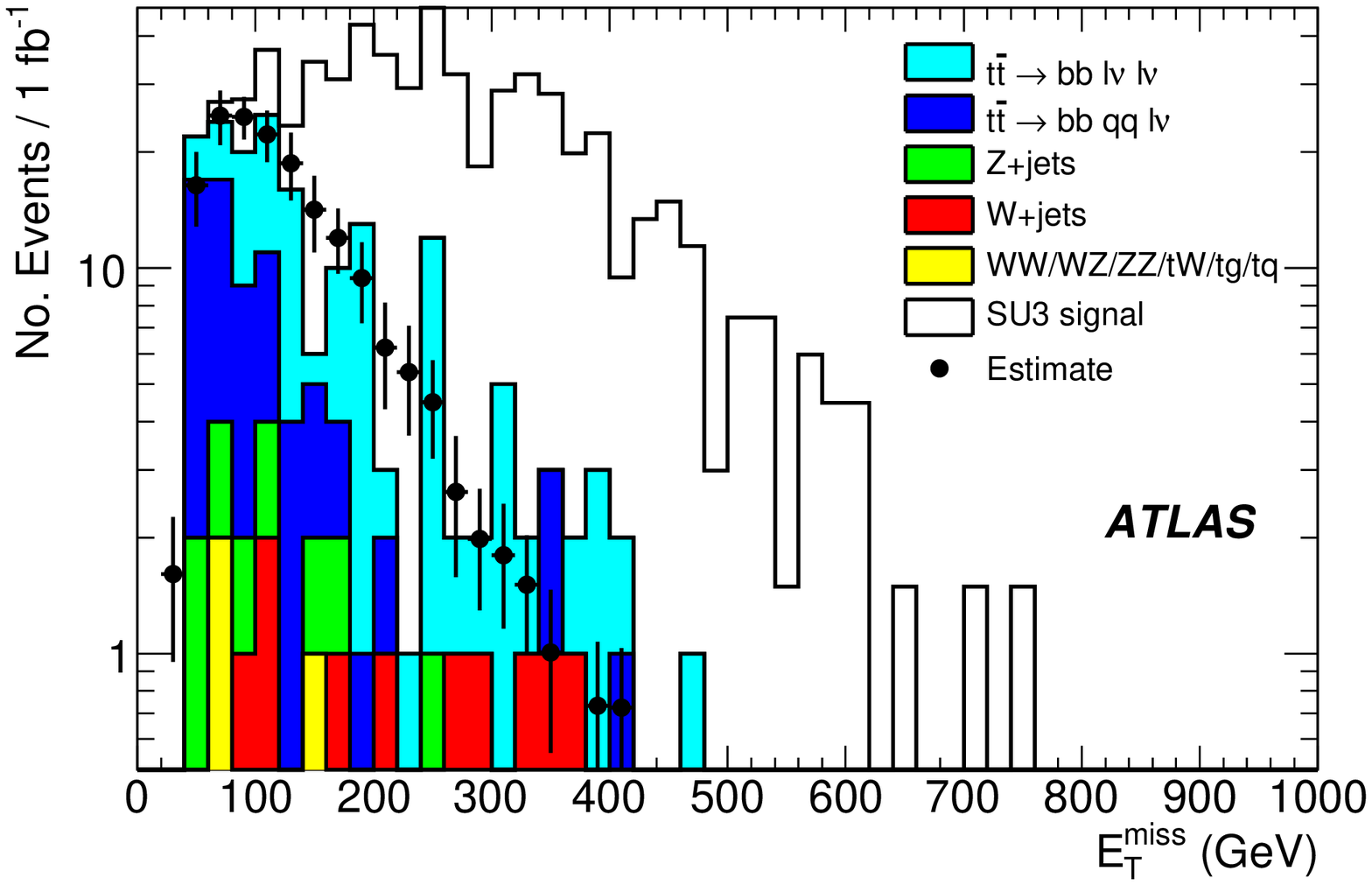} 
\includegraphics[width=7.cm,height=3.5cm]{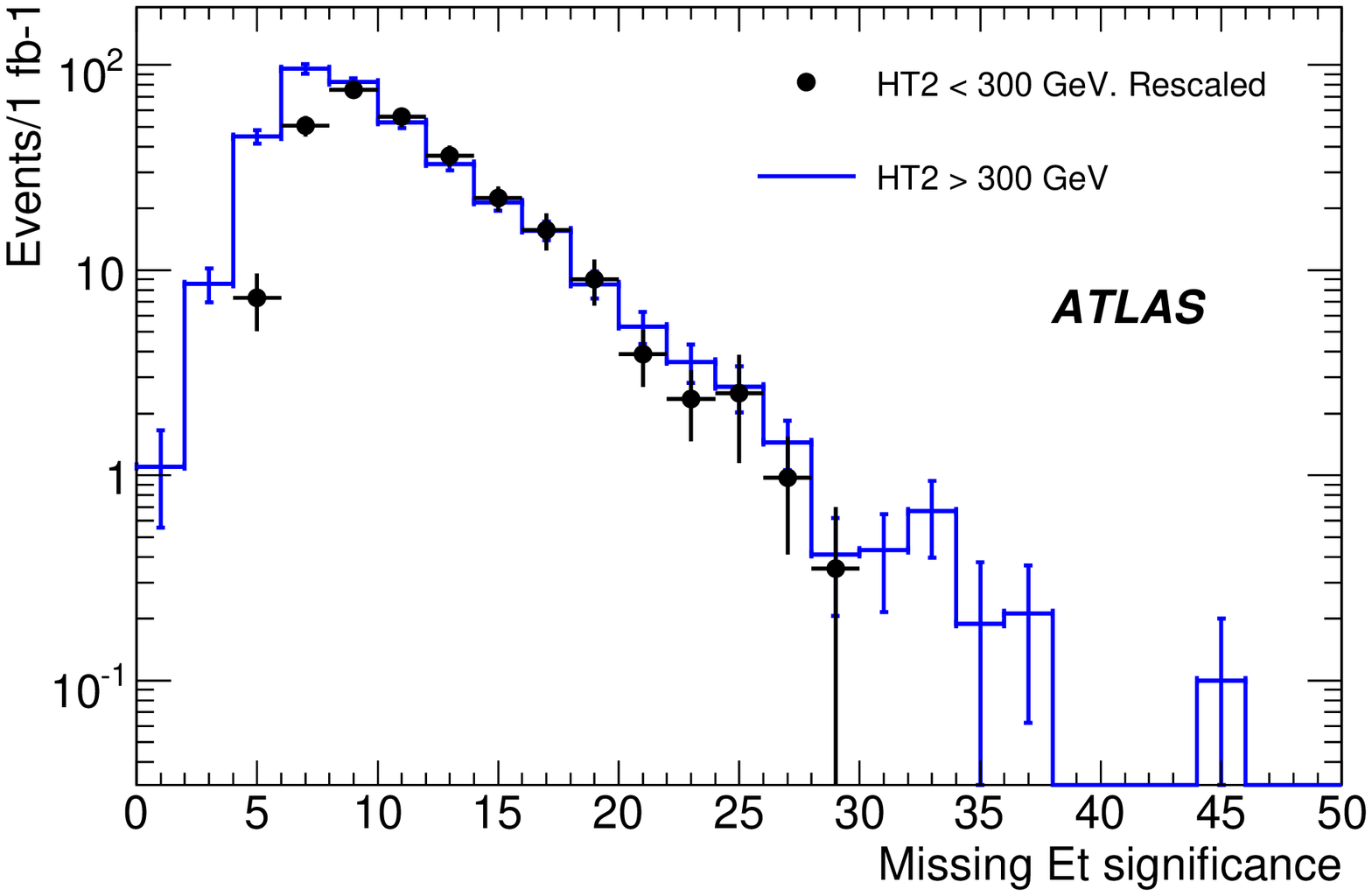} 
\end{center}
\vspace{-0.5cm}
\caption{The estimated and actual $E_{T}^{miss}$ background distributions using the decay resimulation technique without SUSY signal (left). Comparison of actual and predicted $E_{T}^{miss}$ significance distribution using the HT2 technique. }
\end{figure}

\subsection{HT2 TECHNIQUE}
The HT2 variable, which is nearly independent of $E_{T}^{miss}$ , is the scalar $p_{T}$ sum of the second, third and fourth highest $p_{T}$ jets and the reconstructed lepton. It can also be used to estimate the dileptonic background. The $E_{T}^{miss}$ significance (defined as the $E_{T}^{miss}$ divided by $0.49{\times}\sum{E_{T}}$), is estimated from events with low HT2 and normalised to events with high HT2 and low $E_{T}^{miss}$ significance. The right plot in Figure 4 shows a comparison of this estimated distribution and good agreement is seen with the actual distribution. The largest systematic uncertainties on the number of predicted background events in this method are 20$\%$ due to detector uncertainties and $20\%$ due to generator effects.

\vspace{-0.7cm}
\section{NO LEPTON MODE SEARCH}
This search uses the same cuts as the one lepton mode search with a few modifications, described in \cite{susyCSC-ref}. In order to estimate an important background for this mode, $Z$ $\rightarrow$ $\nu\nu$ events, a control sample of $Z$ $\rightarrow$ $ll$ is constructed with a modified no lepton event selection. Two electrons or muons are required, cuts are applied to the $p_{T}$ of the dilepton system instead of the $E_{T}^{miss}$, the invariant mass of the dilepton system is required to be in the range 81 to 101 GeV and the $E_{T}^{miss}$ should be less than 30 GeV. By applying corrects for acceptance, additional kinematic cuts and the lepton ID efficiency the amount of $Z$ $\rightarrow$ $\nu\nu$ background can be estimated. The first two corrections are derived from Monte Carlo and the latter from data using a tag and probe method in a $Z$ $\rightarrow$ eX sample, where X is a loose electron. Good agreement between the estimate and the actual $E_{T}^{miss}$ distribution is shown in Figure 5. The dominant systematics uncertainties on the predicted number of background events come from ALPGEN parameter variation (6.3$\%$) and the soft part of the $E_{T}^{miss}$ scale (4.5$\%$). The total uncertainty, including statistical uncertainties, is 15$\%$.

\begin{figure}[h]
\begin{center}
\includegraphics[width=7.cm,height=3.5cm]{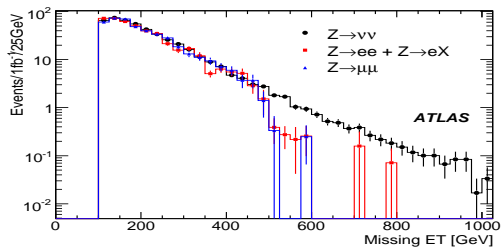} 
\end{center}
\vspace{-0.5cm}
\caption{Comparison of $E_{T}^{miss}$ distribution for $Z$ $\rightarrow$ $\nu\nu$ (derived from $Z$ $\rightarrow$ $ll$), $Z$ $\rightarrow$ $ee$ and $Z$ $\rightarrow$ $\mu\mu$.}
\end{figure}

\vspace{-1cm}
\section{CONCLUSIONS}
A selection of data-driven background methods studied in ATLAS for one and no lepton SUSY search modes has been presented. The sum of top quark pair and $W$ backgrounds in the one lepton mode can be estimated, using the $M_{T}$ method, with an accuracy of 4-8$\%$ statistical uncertainty and 15$\%$ systematic uncertainty. A decay resimulation technique has been developed for use in first data and estimates the dileptonic top quark pair background $E_{T}^{miss}$ distribution with an accuracy of 30$\%$. The fully leptonic background in the one lepton mode search can also be determined with a 5$\%$ statistical uncertainty and 22 $\%$ systematic uncertainty using the HT2 method. The $Z$ $\rightarrow$ $\nu\nu$ background distributions in the no lepton mode search can be estimated with 15$\%$ uncertainty.

\begin{acknowledgments}
I would like to thank the organisers of the ICHEP08 for the invitation to present this poster and the ATLAS collaboration.
\end{acknowledgments}

\end{document}